\documentclass[pre,twocolumn,aps,showpacs]{revtex4}
\usepackage{graphicx}
\usepackage[activeacute,english]{babel}
\begin{document}

\title {   
Tight coupling in thermal Brownian motors}
\author{
A. Gomez-Marin\footnote{email: agomezmarin@gmail.com}
 and J. M. Sancho}
\affiliation{
Facultat de Fisica, Universitat de Barcelona, Diagonal 647, 08028 Barcelona, Spain}

\date{\today}
     
\begin{abstract} 
We study analytically
a thermal Brownian motor model  and calculate exactly  the Onsager coefficients. We show how the reciprocity relation holds and that the determinant of the Onsager matrix vanishes. Such condition implies that the device is built with tight coupling. This explains why Carnot's efficiency can be achieved in the limit of infinitely slow velocities. We also prove that the efficiency at maximum power has the maximum possible value, which corresponds to the Curzon-Alhborn bound. Finally, we discuss the model acting as a Brownian refrigerator. 
\end{abstract}


\maketitle

Can Carnot's efficiency be achieved by a thermal Brownian motor? This question is of great importance, not only for theoretical statistical physicists but for the technological construction of micro and nano machines. In fact, the practical and relevant issue is to investigate  the instrinsic features with which a device should be built in order to operate optimally at the limitations established by first principles.
Here we study a theoretical model for a Brownian motor and show that it has the so called tight coupling property. This gives a fundamental explanation for the fact that the system has Carnot and Curzon-Alhborn's efficiency bounds.

\begin{figure}
\begin{center}
  \includegraphics[ angle=0, width=0.35\textwidth]{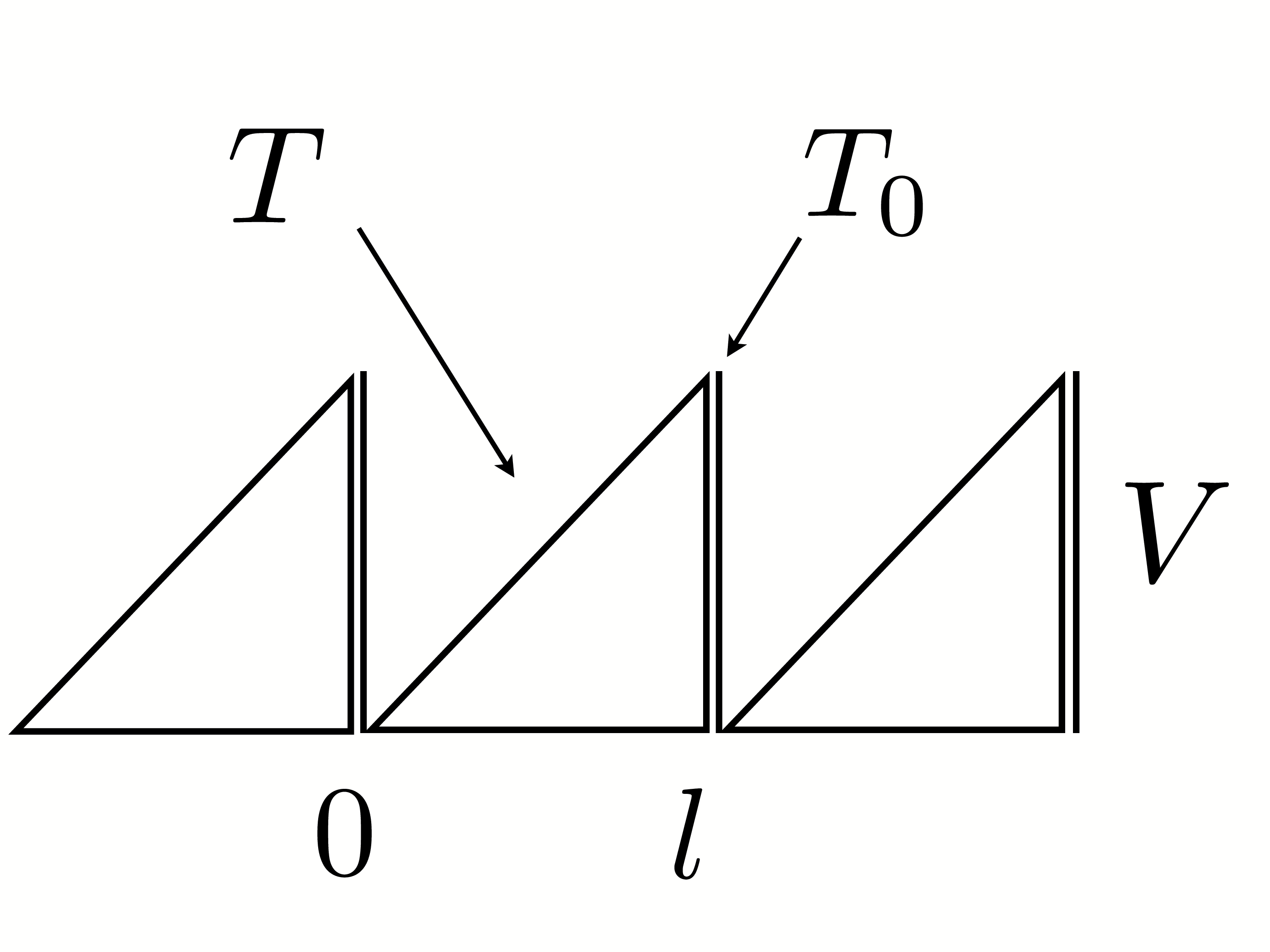}
  \caption{Scheme of the ratchet potential $V(x)$ and temperature profiles in Sakaguchi's model.}
  \label{model}
\end{center}
\end{figure}

Let us consider the model proposed by Sakaguchi \cite{saka} for the Brownian motion of a Feynman-like ratchet device \cite{feyn}. It consists on a Langevin equation in the overdamped regime that accounts only for one degree of freedom at temperature $T$. The particle moves under the action of a spatially periodic and asymmetric potential $V(x)$ and an external force $F$. The equation of motion is
\begin{equation}
\dot{x}=-V'(x)-F+\eta(t),
\label{lang}
\end{equation}
where the friction coefficient has been taken equal to one by redefining the time scale.
The thermal bath is modeled by a Gaussian white noise $\eta(t)$ with the usual fluctuation dissipation relation $<\eta(t)\eta(t')> = 2 T \delta(t-t')$. We  take $k_B=1$, which fixes the energy units. The ratchet potential is illustrated in Fig. (\ref{model}).

The second thermal bath at temperature $T_0$ (which is actually necessary to break detailed balance) is introduced through  the following boundary condition for the steady probability distribution,
\begin{equation}
P(l)=P(0) e^{-V/T_0}.
\label{bc}
\end{equation}
This condition ensures the correct thermodinamical properties of the model and incorporates the transition probabilities suggested by Feynman, namely, that the probability to overcome the barrier in the forward direction (from left to right) is proportional to $e^{-(V+Fl)/T}$ (from the Langevin equation (\ref{lang})), while the probability to cross the barrier backwards is $ e^{-V/T_0}$ (from the boundary condition (\ref{bc})). 
Sakaguchi's model can also be understood as a Brownian particle moving alternately in hot and cold reservoirs along space (see Ref. \cite{beke1}) in the limit in which the cold region tends to zero.

The condition (\ref{bc}) implies that $P(x)$ does not connect with continuity the gap points, in which
$V'(x)$ is infinite. From Eq. (\ref{lang}), one can write the equation for the probability distribution in the steady state,
\begin{equation}
(V'(x)+F)P(x) + T  P'(x) = -\mathcal{J},
\label{flux}
\end{equation}
where $\mathcal{J}$ is the constant and uniform probability  flux. This equation is linear and can be solved analytically with two unknown constants,  $\mathcal{J}$ and $P(0)$, that are evaluated imposing normalization to $P(x)$ and the boundary condition (\ref{bc}). The observable of interest is the mean velocity  and it is  evaluated in terms of the flux, already obtained, as 
\begin{equation} \label{vel}
 v \doteq \langle \dot{x} \rangle = \mathcal{J} l= \frac{a^2 T}{l}\frac{e^{-a}-e^{-b}}{a(e^{-b}-e^{-a})+(1-e^{-b})(1-e^{-a})} 
 \end{equation}
where $a=(V+lF)/T$ and $b=V/T_0$. This expression is exactly the same that the one derived
in Ref. \cite{beke1} if the limit $L_2=0$ is performed.

\begin{figure}
\begin{center}
  \includegraphics[ angle=0, width=0.35\textwidth]{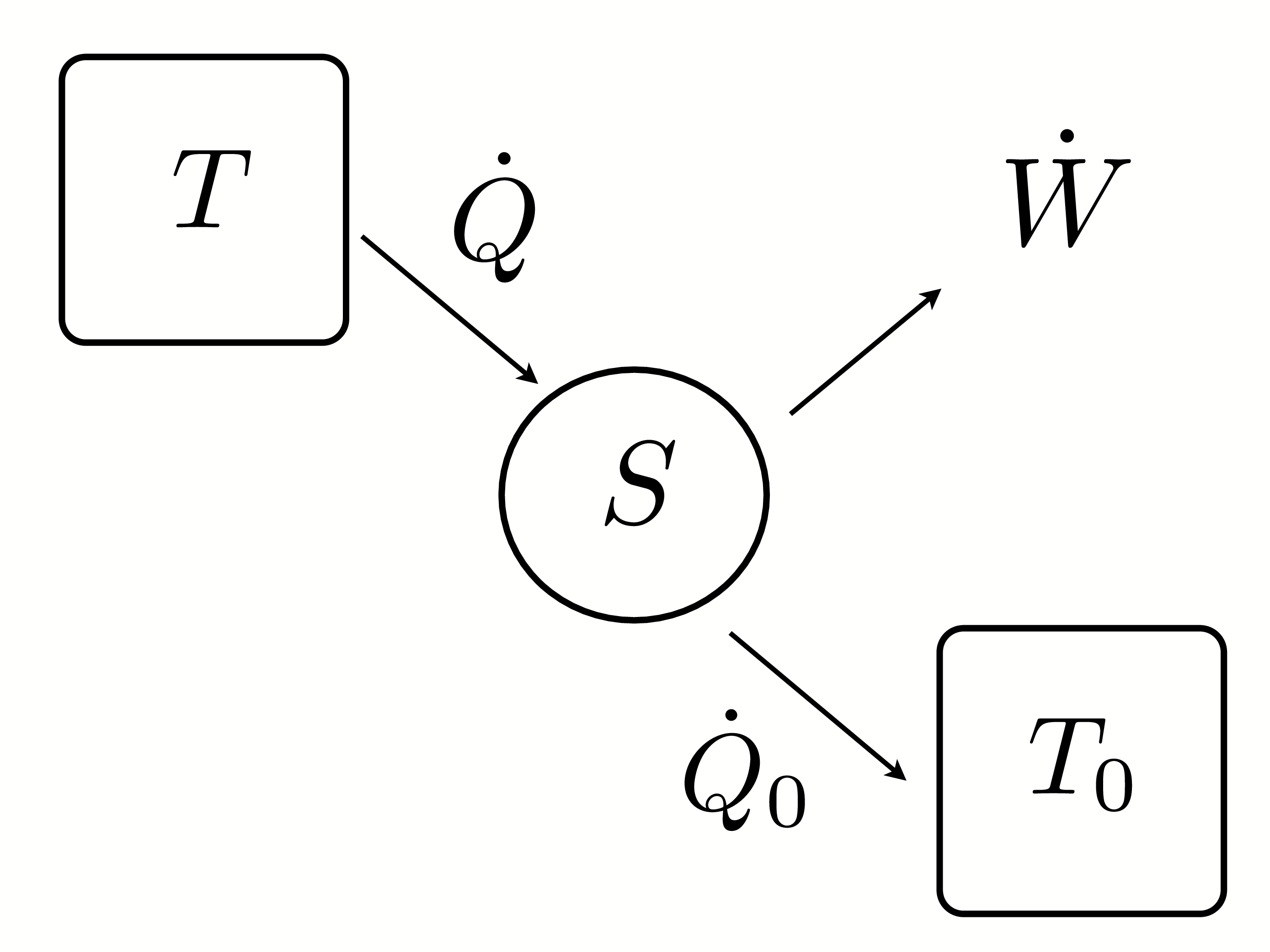}
  \caption{Energetics of a thermal motor:  heat flow $\dot{Q}$ from the hot source at $T$ gets into the system $S$ and mechanical power $\dot{W}$ is performed at the same time that  heat flow $\dot{Q}_0$ goes to the cold source at $T_0$.}
  \label{energy}
\end{center}
\end{figure}

The simplicity of the model allows the  analytical evaluation of other  quantities \cite{saka2}.
For instance, the energetics are straightforward. The mean heat flow released  by the hot reservoir at $T$ is 
 \begin{equation}
 \dot{Q} 
  = (V/l + F )  v, 
\label{Q}
\end{equation}
since the system absorbs it to move against the external load $F$ and potential force $V/l$.
Therefore, the mean heat flow to the cold reservoir is
\begin{equation}
 \dot{Q}_0  = (V/l)  v, 
\label{Q0}
\end{equation}
because every time the potential barrier $V$ is overcome, energy from the hot source is delivered into the bath at $T_0$. Finally, the mean power performed against the conservative force $F$ is just
 \begin{equation}
 \dot{W}  =   F  v.
\end{equation}
Note that the above expressions are consistent with the First Law, 
 $ \dot{Q} = \dot{Q}_{0}+\dot{W} $.  Furthermore, the above characterization is compatible with the one introduced in Ref. \cite{seki}. See Fig. (\ref{energy}) for a scheme of the energetic quantities  involved.

According to the theory of nonequilibrium thermodynamics, the rate of entropy production can be expressed as
\begin{equation} \label{S}
 \dot{S}   =- \frac{    \dot{Q}     }{T}+\frac{   \dot{Q}_0   }{T_0}.
\end{equation}
Now, this expression can be recasted in terms of a second order form involving the thermodynamic forces 
$X_1=F/T$ and $X_2=\Delta T / T^{2}$, where the temperature difference is assumed to be small so that $T-T_0=\Delta T$ and $T \simeq T_0$.
Therefore, in the linear response regime we have 
\begin{equation} \label{ons}
 \dot{S}   =(X_1, X_2)   
  \left( \begin{array}{cc}
L_{11} & L_{12}  \\
L_{21} & L_{22}  \end{array} \right) 
  \left( \begin{array}{c}
  X_1  \\
X_2   \end{array} \right),
\end{equation}
where $L_{ij}$ are the Onsager coefficients.
We take Eq. (\ref{S}), substitute both expressions for the heats (\ref{Q}) and (\ref{Q0}) and expand the velocity $v$ around the equilibrium state  ($F=0$ and $\Delta T=0$) up to first order.  This leads to an identification  of the terms corresponding to every coefficient in Eq. (\ref{ons}).
After some tedious but standard calculations,  one obtains 
\begin{equation} \label{11}
L_{11}= T \left(  \frac{\alpha}  {\sinh (\alpha)} \right)^2,
\end{equation}
\begin{equation} \label{12}
L_{12}= L_{21} = -\frac{VT}{l} \left(  \frac{\alpha}  {\sinh (\alpha)} \right)^2,  
\end{equation}
\begin{equation} \label{22}
L_{22}=\frac{V^2 T}{l^2} \left(  \frac{\alpha}  {\sinh (\alpha)} \right)^2.
\end{equation}
where $\alpha=V/(2T)$.
See \cite{jar} for a similar analysis.
The Onsager coefficients provide a lot of information about the intrinsic nonequilibrium thermodynamic properties of  the system.  First one can check that  the reciprocity relation, 
$L_{12}=L_{21}$, is fulfilled and 
 that the diagonal coefficients 
$L_{11}$ and   
$L_{22}$
are positive as they should.
A relevant relation that is also found is
\begin{equation} \label{det0}
L_{11}L_{22}=L_{12}^{2}.
\end{equation}
This makes the determinant of the $L_{ij}$ matrix equal to zero, which is in agreement with the Second Law. We can operate in a reversible regime and there is no entropy production. In fact, this condition is stronger than just telling us that, in the limit in which $F=0$ and $\Delta T=0$ (motor at rest), no entropy is produced. 
More importantly, the above relation implies that the parameter
\begin{equation}
q \doteq \frac{L_{12}}{\sqrt{L_{11}L_{22}}}=-1,
\end{equation}
which means that the motor is built with the condition of tight coupling \cite{van3}. This is the central  result of this Report.

When writing  the efficiency of a thermal motor to lowest order in $\Delta T /T$, it is found to be a function of the ratio of the thermodynamic forces $\kappa= X_1/X_2$, so that, \cite{van}
\begin{equation}
\eta=\frac{\dot{W}}{\dot{Q}}=\frac{\Delta T}{T} \left( -\kappa\frac{L_{11}}{L_{21}} \right) \frac{\kappa +L_{12}/L_{11}}{\kappa + L_{22}/L_{21}}.
\end{equation}
For the case in which $|q|=1$, the last fraction of the above equation is equal to one. Now, if we stop both fluxes, $J_1=J_2=0$, which would correspond to a vanishing  velocity and  heat flow (but  without the need of setting $F=\Delta T=0$), the above expression reduces  reduces to
\begin{equation} \label{carnot}
\eta=1-\frac{T_0}{T},
\end{equation}
which is the efficiency of Carnot. In general, for  $|q|<1$ it is always below Carnot's. 
The efficiency can also be found using the typical definition 
\begin{equation} \label{eff}
\eta  =\frac{\dot{W}}{\dot{Q}}= \frac{F v}{(V/l+F  )v}=\frac{1}{V/(F l)+1}, 
\end{equation}
and then, by making $v=0$, we obtain from Eq. (\ref{vel}) that 
\begin{equation} \label{v=0}
\frac{T}{T_0}=1+\frac{F l}{V},
\end{equation}
which inserted in Eq. (\ref{eff}), gives again Eq. (\ref{carnot}). 
For non-perfectly tight models, when stopping the power performed (at the stall force), there will still be a leak of heat which will make the efficiency far lower than Carnot's \cite{sbm}.
It is true that achieving such upper bound in the efficiency of the motor in the limit of  zero velocity  
has  little practical relevance. In real devices it is more interesting to study efficiency at  maximum power.
It has been recently proved \cite{van2} that it is given by
\begin{equation}
\eta =1-\left(  \frac{T_0}{T}    \right)^{\rho /2},
\end{equation}
where $\rho = q^2 / (2-q^2)$. Consequently, it is the highest possible just when $|q|=1$, leading to the Curzon-Alhborn bound \cite{ca}
\begin{equation} \label{cal}
\eta= 1-   \sqrt{ \frac{T_0}{T} }.
\end{equation}
The tight coupling of the present model ensures that this bound can be reached too. Let us check it by calculating the efficiency in this regime. Using the relation amongst $F$,  $V$, $l$, $T$ and $T_0$ derived from imposing that $\partial \dot{W} / \partial F=0$ (the function found is hard to write explicitly as a function of the force), one can see that maximum power condition fulfills the  condition
\begin{equation}
\frac{V}{F  l}=\frac{\sqrt{T_0/T}}{1-\sqrt{T_0/T}},
\end{equation}
and thus, when inserted in Eq. (\ref{eff}), leads to (\ref{cal}),
which is precisely  the Curzon-Alhborn efficiency for finite time endoreversible devices.
In this case, one could perform finite time numerical simulations to recover such prediction, as it has been done for a similar model in Ref. \cite{beke1}.

Finally, we want to address the question of converting the Brownian motor into a Brownian refrigerator \cite{van3}. The system is acting as an engine when $\dot{W}>0$ and $\dot{Q}_0>0$. We define the refrigerator mode as $\dot{Q}_0<0$ and $\dot{W}<0$, since the effect of inserting work into the system leads to a heat flux leaving the cold source. Such operation is sometimes regarded as a heat pump \cite{naka}. The existence of a refrigerator mode of operation is easily understood recalling the previous linear irreversible thermodynamics analysis. To put it in words, since a thermal gradient drives the system  against an external mechanical force, due to the cross coupling of the system ($L_{12}$ and $L_{21}$ coefficients), then an external mechanical force can lead to a thermal gradient opposing the one that already exists.
Therefore, any system that can work as an engine, has a region in the parameter space (which may be so small as to be hardy impossible to set) in which will work as a refrigerator.
In Sakaguchi's model, as we can directly see, the reversal of the sign of the heat flowing to the cold reservoir happens simply when the velocity is inverted. Therefore, the transition from the engine mode of operation to the refrigerator occurs when the velocity vanishes, and it is given by Eq. (\ref{v=0}). 
One technical comment to be added is that  the fact that relation (\ref{det0}) is found, leads to no separation between the engine and refrigerator areas \cite{jar}. Obviously, the tight coupling also ensures an optimal performance of the refrigerator mode of operation.

In conclusion, we have studied a simple model of thermal engine described by a Langevin equation which allows analytical calculations of  relevant thermodynamic quantities. We give a fundamental explanation for the fact that the device can achieve optimal energetic performances.
The question of whether a similar tight coupled construction is possible for a system in the underdamped regime is of crucial importance for the applicability to real microscopic machines, since heat flow driven by kinetic energy is relevant when inertia is taken into account \cite{astu,seki2,ai}.

Fruitful discussions with Prof. Chris Van den Broeck are gratefully acknowledged.
This work was finished during a kind invitation stay by Prof. Katja Lindenberg at the University of California San Diego.
This research was supported by Ministerio de Educaci\'on y Ciencia (Spain)
under the  project FIS2006-11452-C03-01 and the grant  FPU-AP-2004-0770.

\end{document}